**Vibrotactile versus Visual Stimulation in Learning the Piano**

Matteo A. Coscia[1], Mazen Al Borno[2]

University of Colorado Denver | Anschutz Medical Campus

1: matteo.coscia@colorado.edu,  2: mazen.alborno@ucdenver.edu (corresponding author)

Department of Computer Science and Engineering

Denver, Colorado, United-States

**Abstract**

Vibrotactile stimulation has been explored to accelerate the acquisition of motor skills involving finger movements (Gemicioglu et al. 2022, Markow et al. 2010, Seim et al. 2017). This study evaluates the effectiveness of vibrotactile stimulation compared to visual feedback in learning a 14-note one-handed tune on the piano. In the experiment, 14 subjects with no prior piano experience were exposed to both vibrotactile and visual stimulation to determine which was more effective. Subjects were randomized 1:1 in a group that first receives vibrotactile stimulation, then visual stimulation or in a group that first receives visual stimulation, then vibrotactile stimulation. Effectiveness was measured by evaluating the timing error and accuracy. Results from our study indicated that the timing error for vibrotactile stimulation was 12.1% (SD 6.0%), while the equivalent for visual stimulation was 22.3% (SD 10.3%). The accuracy for vibrotactile stimulation was 69.2% (SD 27.2%), while the equivalent for visual stimulation was 91.3% (SD 13.5%). It was observed that vibrotactile stimulation was generally more effective at minimizing the timing error at which the notes were hit compared to visual stimulation, and no statistically significant differences were found in accuracy ($U = 61$, $p = 0.0930$).

Research Highlights:

- Vibrotactile stimulation reduces timing errors in piano learning compared to visual stimulation.

- No significant difference in accuracy between vibrotactile and visual stimulation.

- Study highlights potential applications of vibrotactile stimulation in passive motor learning.

## Introduction

Vibrotactile stimulation is a mechanical stimulation that is produced with actuators placed on the skin, producing a similar sensation as a phone vibrating (Seim et al. 2015). Previous studies have shown that vibrotactile stimulation can accelerate the acquisition of motor skills involving finger movements such as keyboard typing and playing the piano (Gemicioglu et al. 2022, Markow et al. 2010, Seim et al. 2017); however, none have compared the effectiveness of vibrotactile stimulation with visual stimulation. Markow et al. 2010 compared active learning and passive learning through vibrotactile stimulation in helping subjects learn how to play the piano. Passive learning occurs when a subject learns a skill without actively paying attention to the task or practicing it (e.g., while engaged in another activity). Active learning is when the subject undergoes repeated and engaged practice to learn the skill. The researchers implemented the passive training using a haptic glove that provided stimulation to the user. They found no significant differences in performance after the use of the haptic glove compared to before the use of the haptic glove. Gemicioglu et al. 2022 discussed how sessions of active piano learning can be interleaved with passive learning sessions to accelerate piano skill acquisition. However, there is a gap in the field in comparing the two methods of visual and vibrotactile stimulation. In this paper, we investigate the relative effectiveness of visual stimulation (delivered in a passive learning style) and vibrotactile stimulation (delivered in a passive learning style) in helping subjects learn to play a 14-note one-handed tune on the piano (i.e., a sequence of musical notes arranged in a specific pattern).

By passing vibrations to different fingers using vibrating motors, vibrotactile stimulation can help subjects passively learn the sequence in which the motors vibrate (Fig. 1(a); Seim et al. 2015). In our work, visual stimulation involves the subject watching lights flash on a physical keyboard, indicating which notes to press (Fig. 1(b)). With visual stimulation, the subjects can see the notes to play, but they do not have any indication of which fingers to use. In this work, we inquired on how the effectiveness of vibrotactile stimulation and visual stimulation for learning the piano compared. This was investigated through an experiment with subjects without prior piano experience, with the hypothesis that vibrotactile stimulation is more effective than visual stimulation at helping people learn a short one-handed tune on the piano.

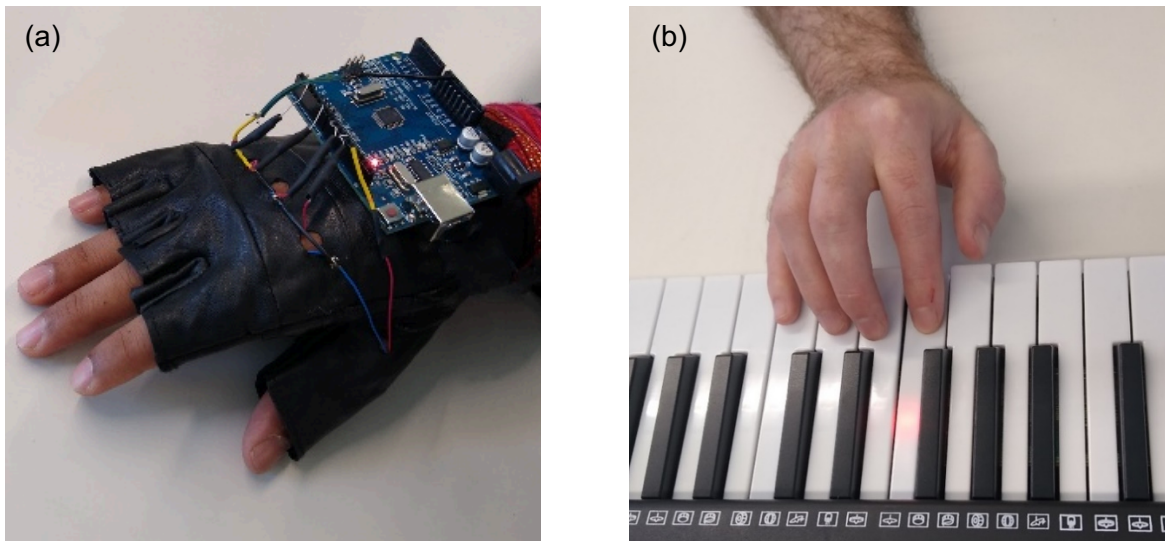

**Figure 1.** (a) Vibrotactile stimulation glove that stimulate the knuckles. (b) Subject using keyboard lights to learn a tune.

# Related Work

Vibrotactile stimulation has applications in virtual reality (Oagaz et al. 2021), motor learning (Markow et al. 2010), and non-invasive treatments of motor disorders such as Parkinson's disease (Pfeifer et al. 2021). In this work, we are interested in vibrotactile stimulation for learning sequences of finger movements. Similarly to previous work, we place the vibrating motors at the back of the fingers, near the knuckles (Seim et al. 2015). We prefer this location instead of the fingertips because the later would inhibit the subject's ability to manipulate objects, including the keys on the piano. Hsiao et al. 2011 developed a prototype to potentially improve piano lessons with vibrotactile stimulation. In their setup, both the piano teacher and the student wore vibrotactile stimulation gloves. The glove for the teacher recorded the finger tapping and hand movements, which were immediately signaled to a microcontroller and vibrated into the student's glove. In this way, the teacher would not have to physically instruct the student on where to place their fingers to play certain notes: they can learn because they feel the vibrations on their fingers rather than having to look at what the teacher is doing. In our work, the vibrations applied on the fingers are derived from the Audacity library (Audacity Team 2022) to generate an idealized representation of a performance (i.e., that determines which key to press, when, and for how long).

Vibrotactile feedback has been explored in areas aside from finger motions. Mihail et al. 2023 found that users have a favorable perception of vibrotactile feedback on large displays, with a preference for feedback on the fingers compared to the wrist and forearm. Other studies found that visual and vibrotactile feedback have similar performances to correct for incorrect body postures in upper limb robot-assisted rehabilitation and ergonomics (Scotto di Luzio et al.

2019; Zheng et al. 2012), but vibrotactile feedback may result in faster reaction times (Forster et al. 2002).

Other previous research in this area has discussed the effectiveness of vibrotactile stimulation to learn a sequence of finger movements. Huang et al. 2008 introduced the PianoTouch, a wearable haptic piano instruction system for the passive learning of piano skills. They conducted experiments on four subjects, testing the efficacy of vibrotactile stimulation for learning the tunes "Amazing Grace" and "Jingle Bells" on the piano. They found that vibrotactile stimulation was effective at teaching these songs passively: with vibrotactile stimulation, subjects hit 0 - 3 incorrect notes per song, and with active practice, subjects hit 4 – 7 incorrect notes per song. Others have used vibrotactile stimulation to teach subjects a randomized numeric keypad (Seim et al. 2017).  In these studies, vibrotactile stimulation was shown to be the most effective out of the other options tested (i.e., audio and active practice). However, none of these studies have compared vibrotactile stimulation directly to visual stimulation, both in the passive form. This study expands on the existing work in the field by making this direct comparison.

# Methods

Our study used the following materials: a keyboard with lights embedded in the keys, a vibrotactile stimulation glove, a microphone for recording the subjects playing the tune and Audacity's tone generator (software for generating the ideal tune). The vibrotactile stimulation glove consists of a microcontroller from Arduino (Uno R3) and 5 eccentric rotating mass (ERM) vibration motors from Seeed Technology (Model # 316040001). The motors were placed on the proximal phalange of each finger to ensure maximal differentiation as to which finger is receiving the vibration. The vibrational signal was sent to the motor of choice for a specified duration (250 ms) and a voltage of 3.3 V (corresponding to a 195-200 Hz frequency) using Arduino C++. We conducted a power analysis based on a 2-sample t-test with 2-sided α level set at 0.05 and power at 80% for the ability to detect a 7% difference in timing error between vibrotactile and visual feedback. The analysis resulted in having 7 subjects per group; therefore, 14 subjects in total were recruited within our institution for the experiment. Subjects were randomized 1:1 into two groups, one that first receives visual stimulation, then vibrotactile stimulation, and the second group that first receives vibrotactile stimulation, then visual stimulation. The subjects were healthy and we had 9 males and 5 females with an average age of 32 (SD 10.7) years. We did not expect the gender of the subjects to have an impact on the results, as no prior studies have reported notable differences. Differences between age groups may be present and would be interesting to investigate in future work. We recruited subjects that had no prior piano experience as we were interested in investigating the effects of visual and vibrotactile feedback in motor learning in novice subjects. The total time for the experiments was about 45 minutes per subject. Subjects understood and consented to the protocol approved by the Institutional Review Board of the University of Colorado Anschutz Medical Campus.

We conducted our experiments for both groups with the same tune, a segment chosen from a song titled "Morning" from a beginner piano lesson book (Faber, N., & Faber, R., 1998; see Figure 1 in the supplementary material). We identified a 90-bpm tune that could be truncated to 14 notes, be played with the right hand only and did not require the hand to move across the keyboard. The tune chosen was representative of simple piano tunes using white notes only, so the results might be expected to hold for other tunes of this kind, as long as they are approximately the same size and do not require the movement of the hands across the keyboard. Our experiment used 20 minutes of passive vibrotactile stimulation to teach the subjects the sequence of finger movements. We selected this based on prior work finding that 20 minutes of passive vibrotactile stimulation was effective in teaching people to play a song on the piano (Kohlsdorf & Starner 2010). The participants did not wear headphones. They were also not blindfolded during the duration of the study. By using Audacity tones, we programmed the way the tune should ideally be played by specifying the parameters of the sounds in the program (i.e., frequency, amplitude and duration). Each note was programmed with its specific frequency, onset time, and duration, determined with the generate tone function in Audacity.

In the first group, the subjects watched lights on a keyboard flash to the same tune for 3 minutes (ten times). This showed them where they were supposed to press and for what amount of time to press, but not where to place all their fingers. After the 3 minutes, they were instructed to play the song again and were given no verbal instructions on what fingers to use or what the starting note was, and the song was recorded. Then, the subjects wore the glove for 20 minutes (67 repeats of the tune) of passive vibrotactile stimulation, where the fingers they should use to press the key were stimulated. The duration of the vibration was not set proportionally to the duration the key needed to be pressed. Therefore, we would not expect the vibrotactile stimulation to improve the accuracy of the duration a note is pressed, but this could be interesting to examine in a follow-up study.

After those 20 minutes, the subjects removed the glove and were told "your pinky starts here, on G," and that each of their fingers should be on a white key, with their thumb on C, pointer on D, middle finger on E, and so on. They were then instructed to play the song, and it was recorded. In the second group, the same procedure was repeated with the same tune, except that the vibrotactile stimulation was applied first, then the visual stimulation.

At the end of the experiments, the recordings were compared to the ideal spectrogram using Audacity by taking the timing bar and placing it at the start of the note in the recorded version of the tune. These values were compared to the ideal timing and note sequence using three different metrics. The first metric was the average timing error percentage of the notes played, which described the difference between the time the subject was supposed to hit a note and the time the subject actually hit a note, as a percentage. For example, if the subject hit the note at 3.45 s instead of 3.0 s, the timing error would be 15%. The second metric was the accuracy of the notes played, which refers to the percentage of notes the subject played in the correct order. When the note was played and either the note before or after it was correct, it was counted as an accurate hit. When the note was omitted or not in the correct position, it was counted as inaccurate. The third metric was the 2-norm difference, which was calculated by averaging the norm of the difference between the two values, i.e., $\| X - Y \|_2$, where X is the expected timing and Y is the actual timing the note was hit.

## Results

We will now compare how the subjects learned to play the tune after receiving visual or vibrotactile stimulation. In Fig. 2(a), we plot the average timing error percentage for subjects 1 to 7 when visual stimulation is applied first, followed by vibrotactile stimulation (first group). We observe that for 5 of the 7 subjects, the vibrotactile stimulation reduced the timing errors. In Fig. 2(b), we plot the average timing error percentage for subjects 8 to 14 when vibrotactile stimulation is applied first, followed by visual stimulation (second group). We observe that for 6 of the 7 subjects, the visual stimulation increased the percentage error in timing.

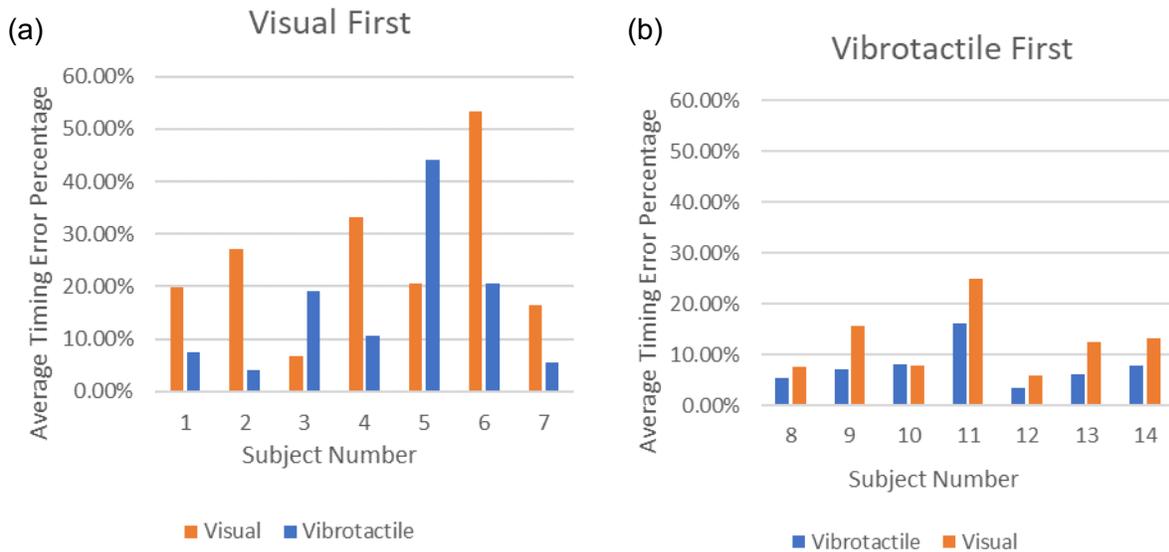

**Figure 2.** (a) The average timing error percentage for subjects 1 to 7 when the visual stimulation is applied first, followed by the vibrotactile stimulation. (b) The average timing error percentage for subjects 8 to 14 when the vibrotactile stimulation is applied first, followed by the visual stimulation.

In Fig. 3(a), we compare the average timing errors percentages after vibrotactile stimulation and after visual stimulation, but this time averaged over all subjects regardless of order of presentation. We observe that the average timing error with vibrotactile stimulation is 12.6% (SD 11.5%) and with visual stimulation is 19.9% (SD 13.7%). A Shapiro-Wilk test indicated that the vibrotactile stimulation dataset was not normally distributed (W = 0.623; p <

0.001); therefore, the Mann-Whitney U-test was performed to compare the visual and vibrotactile stimulation groups. The difference between the average timing error with vibrotactile stimulation and that with visual stimulation was statistically significant (U = 54, p = 0.0455). In Fig. 3(b), we compare the average accuracy percentages after vibrotactile stimulation and after visual stimulation. We did not find a statistically significant difference in accuracy between visual and vibrotactile stimulation (U = 61, p = 0.0930).

(a)

(b)

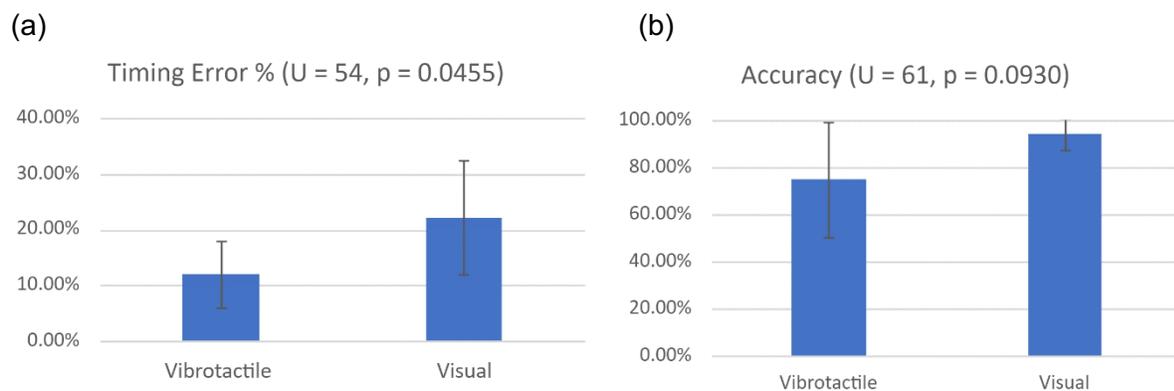

**Figure 3.** (a) Average timing error percentage after vibrotactile stimulation and visual stimulation, averaged over all subjects, regardless of order of presentation. (b) Average accuracy percentage after vibrotactile stimulation and visual stimulation. Error bars represent on standard deviation in the timing and accuracy errors.

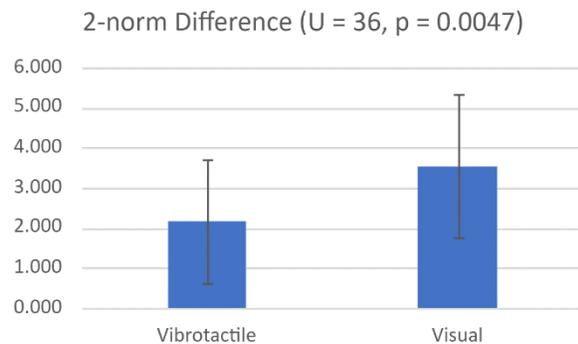

**Figure 4**. 2-norm difference for the timing of notes metric after vibrotactile stimulation and visual stimulation. Error bars represent standard deviation in the 2-norm difference errors.

In Fig. 4, we observe that the average 2-norm difference was larger for visual stimulation compared to vibrotactile stimulation. The data was found to be statistically significant (U = 36, p

= 0.0047). We also investigated the Pearson cross-correlation coefficient metric, in which no statistically significant differences between groups were found.

**Discussion**

In this study, we found that vibrotactile stimulation was more effective at improving the timing of the notes played, indicating that learning motor sequences could be easier with tactile feedback rather than visual feedback, perhaps due to the faster reaction times (Akamatsu et al. 1995; Westebring-van der Putten et al. 2009) or more robust motor memories (Amoruso et al. 2021) of the former. In practical settings, this indicates that vibrotactile stimulation combined with traditional piano books could help students learn to associate notes on paper with notes on the piano. When the vibrotactile stimulation was done first, the visual stimulation had slightly larger timing errors than the previous vibrotactile stimulation. When the visual stimulation was done first, the subsequent vibrotactile stimulation had smaller timing errors than the previous visual stimulation. Therefore, vibrotactile stimulation was more effective than visual stimulation in terms of timing accuracy, regardless of the sequence in which the participants were presented with the different stimulations. Although we did not collect data on the subjects' subjective experience with the vibrotactile stimulation, subjects did not informally report the stimulation to be annoying or unpleasant. One limitation of our experiment was that during the vibrotactile stimulation, some of the subjects were more focused on the vibrations, while others were more focused on talking, looking at their phones, or another activity. Hence, we did not ensure that all our subjects were uniformly focusing on the vibrations as a motivation for passive learning with vibrotactile feedback is that motor learning could still occur even when the subjects are not actively focusing on the task. An open question for future work would be to determine the effects of distraction on motor acquisition with vibrotactile feedback. This could be assessed rigorously by measuring how much time the subjects looked at the piano keys or the vibrotactile stimulation system with eye tracking and having a distraction task.

Our comparison between visual and vibrotactile feedback involve different durations, that is, 3 minutes and 20 minutes, respectively. We chose to only have 3 minutes of visual feedback because it is an active learning process which could be tiring for subjects for longer durations. It is also not clear whether 20 minutes of vibrotactile stimulation was necessary or whether shorter durations would have been equally effective for learning the sequence of finger movements. In future work, it would be interesting to explore the interplay between active practice and the type of feedback as it could differentially impact motor memory during learning. It is not currently known whether vibrotactile stimulation can help learn complex multi-joint movements (in addition to finger sequences) as required in motor skills such as martial arts or playing the violin. An open problem in the field is to determine whether the stimulation is more effective in teaching kinematic or muscle activity patterns. In future work, we can build on the system to include the movement of the hand across the piano, possibly conveyed by vibration motors on the wrist or forearm. This would open the possibilities to learning many more tunes. We could also combine the glove with visual stimulation so the user can see where their hand should be on the piano and receive the information about which finger to press through the glove.

## Data Availability Statement

The dataset supporting the conclusions of this article is available on

https://github.com/MatteoCoscia/Vibrotactile-Stimulation/blob/main/Data.xlsx. The software for

the stimulation is available on https://github.com/MatteoCoscia/Vibrotactile-Stimulation.

## Funding

This work was supported in part by the NIH/NCATS Colorado CTSA Grant Number UL1

TR002535.

## Acknowledgements

Not Applicable.

## Competing Interests

The authors have no competing interests.

## Contributions

All authors were responsible for the conceptualization of the project. The first author conducted

the human subjects experiments, developed the software, conducted the analysis and wrote the

manuscript. The second author helped conduct the analysis, edit the manuscript and supervise all

aspects of the project.